\documentclass[epj]{webofc}
\usepackage[varg]{txfonts}   
\wocname{EPJ Web of Conferences}
\woctitle{ICNFP 2016}
\begin{document}
\selectlanguage{english}
\title{$CP$ violation in $D$ meson decays at Belle}
%
%

\author{K. Prasanth \inst{1}\fnsep\thanks{\email{prasanthkp@physics.iitm.ac.in}} \\$(${\it On behalf of Belle collaboration}$)$
}

\institute{Indian Institute of Technology Madras, Chennai $-$ 600036 
}

\abstract{%
  $CP$ violation and branching fraction measurements in $D$ decays are interesting topics as any difference with respect to the Standard Model prediction would be an indication of new physics. With the large data
sample collected by the Belle detector, which sits at the interaction point of KEKB asymmetric $e^{+} e^{-}$
collider in Japan, we present the results of searches for $CP$ violation in $D^{0}\rightarrow V\gamma~(V = \phi, \overline{K}^{*0}, \rho^{0})$ decay and the rare $D$ decay $D^{0}\rightarrow \gamma \gamma$.
}
\maketitle
%
\section{Introduction}
\label{intro}
~~~$CP$ violation in the charm sector is an interesting topic as it is very sensitive to the new physics. This is because Standard Model $CP$ violation in the $D$ meson system is very small $\mathcal{O}(10^{-3})$. Therefore even a small enhancement in $CP$ asymmetry will be due to new physics. Any enhancement comes from loop-level contributions from non-Standard Model particles or interactions. Here we present the preliminary branching fraction $(\mathcal{B})$ and $CP$ asymmetry $(A_{CP})$ results for $D^{0}\rightarrow V\gamma~(V = \phi, \overline{K}^{*0}, \rho^{0})$ decay and the measurements of branching fraction $(\mathcal{B})$ of $D^{0} \rightarrow \gamma \gamma$ decay with the Belle data set corresponding to an integrated luminosity of 943 fb$^{-1}$ and 832 fb$^{-1}$, respectively.
\section{$D^{0}\rightarrow V\gamma$ decay}
\label{sec-1}
~~~The decay $D^{0}\rightarrow V\gamma~(V = \phi, \overline{K}^{*0}, \rho^{0})$ is sensitive to new physics via $A_{CP}$ measurements. This is because of the contributions from the chromomagnetic dipole operators [1], [2] that can enhance $A_{CP}$ from the expected value of zero. So far no $A_{CP}$ measurements have been reported for this mode. The previous $\mathcal{B}$ measurements [3], [4] of $D^{0}\rightarrow V\gamma$ mode are shown in Table 1. With the 943 fb$^{-1}$ Belle data set, which is very much larger than the one used in the previous measurements, we expect to get a more precise measurement of $\mathcal{B}$ for this decay channel. This is the first analysis to measure the $\mathcal{B}$ for $D^{0}\rightarrow \rho^{0} \gamma$ mode.

\begin{table}[ht!]
\centering
\caption{Previous $\mathcal{B}$ measurements in $D^{0}\rightarrow V \gamma$ decays.}
\begin{tabular}{llll}
  \hline 

Collaboration	&	Luminosity & Decay mode	& Branching fraction $\left(\mathcal{B}\right)$ \\	
    \hline
~~Belle$^{\bf[3]}$	&	~~78 fb$^{-1}$ & $D^{0}\rightarrow \phi \gamma$	& $\left( 2.60^{+0.70}_{-0.61} (\mathrm{stat}) ^{+0.15}_{-0.17} (\mathrm{syst}) \right) \times 10^{-5}$\\
BABAR$^{\bf[4]}$	& 	387 fb$^{-1}$ & $D^{0}\rightarrow \phi \gamma$	&  $\left(2.73 \pm 0.30 (\mathrm{stat}) \pm 0.26 (\mathrm{syst}) \right) \times 10^{-5}$ \\
BABAR$^{\bf[4]}$	&   387 fb$^{-1}$ & $D^{0}\rightarrow \overline{K}^{*0} \gamma$	& $\left(3.22 \pm 0.20 (\mathrm{stat})  \pm 0.27 (\mathrm{syst}) \right) \times 10^{-4}$ \\
  
   \hline	
  \end{tabular}

  \end{table}

\subsection{Selection criteria, fitting and signal extraction}
\label{sec-2}
~~~Selection criteria are defined using a Monte Carlo (MC) simulation study performed on data produced by the {\tt EVTGEN} [5] and {\tt GEANT3} [6] packages; the former includes the effects of final state radiations (FSR). The $D$ mesons are required to come from the $D^{*\pm}$ meson by $D^{*\pm}\rightarrow D^{0} \pi_{\mathrm{slow}}^{\pm}$ decay. As the final state is the same for both $D^{0}$ and its conjugate $\overline{D^{0}}$ decay, we use the charge of $\pi_{\mathrm{slow}}^{\pm}$ to distinguish the flavour of the $D$ meson. The $\pi_{\mathrm{slow}}$ is so called because it carries very little momentum compared to the $D$ meson. The vector mesons are reconstructed from the following decay channels: $\phi\rightarrow K^{+} K^{-}$, $\overline{K}^{*0}\rightarrow K^{-} \pi^{+}$ and $\rho^{0}\rightarrow \pi^{+} \pi^{-}$. The selection criteria for the variables are chosen to maximize the significance, defined as $\frac{N_{\mathrm{sig}}}{\sqrt{N_{\mathrm{sig}} + N_{\mathrm{bkg}}}}$, where $N_{\mathrm{sig}}$ and $N_{\mathrm{bkg}}$ are the number of signal and background events in the defined signal region. A tight mass window of 11 MeV/c$^{2}$ is applied for the $\phi$ candidates around its nominal mass [7] due to the narrow resonance. The mass window for the $\overline{K}^{*0}$ and $\rho^{0}$ candidates are 60 MeV/c$^{2}$ and 150 MeV/c$^{2}$ respectively. The photon candidates are selected with energy greater than 540 MeV. In order to suppress the merged photon cluster, we made a cut on the ratio of 3 $\times$ 3 array ($E_{9}$) to 5 $\times$ 5 array ($E_{25}$) of Electromagnetic Calorimeter (ECL) crystals to be greater than 0.94. A vertex fit is performed for $D^{*}$ and $D$ mesons where the candidates with confidence level less than 10$^{-3}$ are rejected. The $D^{*}$ meson fit also has an Interaction-Point (IP) constraint, which ensures that the $D^{*}$ daughter particles  originate from the $e^{+}e^{-}$ interaction region. We defined a variable $ q \equiv M(D^{*+}) - M(D^{0}) - M(\pi^{+})$, where $M(D^{*+})$, $M(D^{0})$, $M(\pi^{+})$ are the masses of $D^{*}$ meson, $D$ meson and $\pi$ meson, respectively, which is the total energy released in a $D^{*}$ decay. The $D^{*}$ candidates with a $q$ value within 0.6 MeV/c$^{2}$ of the nominal value [7] are selected. A cut on the momentum of $D^{*}$ meson in the center-of-mass (CMS) system is also applied for the modes, which are 2.42 GeV/c for the $\phi$ mode, 2.17 GeV/c for the $\overline{K}^{*0}$ mode and 2.72 GeV/c for the $\rho^{0}$ mode.

The measurements of $\mathcal{B}$ and $A_{CP}$ are performed by using control samples for all the three modes. This procedure ease the systematic studies as many common effects get cancelled. These control samples are $D^{0}\rightarrow K^{+} K^{-}$ for the $\phi$ mode, $D^{0}\rightarrow K^{-} \pi^{+}$ for the $\overline{K}^{*0}$ mode and $D^{0}\rightarrow \pi^{+} \pi^{-}$ for the $\rho^{0}$ mode. The branching fraction is calculated as

\begin{equation}
			\mathcal{B}_{\mathrm{sig}} = \mathcal{B}_{\mathrm{norm}} \times \frac{N_{\mathrm{sig}}}{N_{\mathrm{norm}}} \times \frac{\epsilon_{\mathrm{norm}}}{\epsilon_{\mathrm{sig}}},
\end{equation}
where $N_{\mathrm{sig}}$ $(N_{\mathrm{norm}})$, $\mathcal{B}_{\mathrm{sig}}$ $(\mathcal{B}_{\mathrm{norm}})$ and $\epsilon_{\mathrm{sig}}$ $(\epsilon_{\mathrm{norm}})$ are the yields from the fit, branching fractions and reconstruction efficiencies for the signal (control sample) modes. 

The $CP$ asymmetry can be derived from the following equation
\begin{equation}
A_{\mathrm{raw}} = A_{CP} + A_{FB} + A_{\epsilon}^{\pm},
\end{equation}
where $A_{\mathrm{raw}} = \frac{N(D^{0}) - N(\overline{D^{0}})} {N(D^{0}) + N(\overline{D^{0}})}$ is the raw asymmetry ($N(D^{0})$, $N(\overline{D^{0}})$ are the yields of $D^{0}$ and $\overline{D^{0}}$ from the fit to the data), $A_{FB}$ [7] is the asymmetry in the production of $D^{*+}$ and $D^{*-}$ due to the interference between $\gamma^{*}$ and $Z^{0}$ bosons in the $e^{+}e^{-}\rightarrow c\overline{c}$ process and $A_{\epsilon}^{\pm}$ is the asymmetry in the efficiency of the reconstruction of positive and negative charged particles, respectively. Thus $A_{CP}$ can be extracted by using the information from the control samples in the respective signal modes. Since $A_{FB}$ and $A_{\epsilon}^{\pm}$ is common to both signal and control sample modes, they are cancelled out. Finally, 
\begin{equation}
A_{CP}^{\mathrm{sig}} =A_{\mathrm{raw}}^{\mathrm{sig}} -A_{\mathrm{raw}}^{\mathrm{norm}} + A_{CP}^{\mathrm{norm}},
\end{equation}
where subscript sig (norm) corresponds to the signal (control) sample.

~~~The dominant background for the signal mode comes from the decay of a $\pi^{0}$ meson to a pair of photons. Here one of the the photons from the $\pi^{0}$ meson is misreconstructed as the signal candidate. To reduce these events, a $\pi^{0}$ mass veto is applied with a neural network variable [8], [9] obtained from the two mass veto variables. The signal photon is combined with all other photons in an event with an energy 30 (75) MeV and when the diphoton invariant mass is close to the nominal mass of $\pi^{0}$ meson, the pair is fed to the neural network. A selection on the output of the neural network results in the retention of  85$\%$ of the signal while rejecting 60$\%$ of the background. 

~~~To extract the signal yield and then $A_{CP}$, a simultaneous two-dimensional unbinned extended maximum likelihood fit is performed with the variables $M_{D^{0}}$ and $\cos{\theta_{H}}$, where $\theta_{H}$ is the the angle between one of the $V$ daughter particles with $D$ meson in the $V$ rest frame. The signal $\cos{\theta_{H}}$ distribution is expected to have the form 1$-\cos^{2}{\theta_{H}}$ due to the conservation of angular momentum while the background distributions do not. The range for $M_{D^{0}}$ is 1.67 $< M_{D^{0}} <$ 2.06 GeV/c$^{2}$ for all the three modes. A tighter cut of $-$0.8 $< \cos{\theta_{H}} <$ 0.4 is also applied in $\overline{K}^{*0}$ and $\rho^{0}$ modes to suppress certain peaking backgrounds. The signal reconstruction efficiency is estimated from signal MC to be 9.7$\%$, 7.8$\%$ and 6.8$\%$ for $\phi$, $\overline{K}^{*}$ and $\rho^{0}$ modes, respectively. 

~~~The invariant mass distribution of signal events in $\phi$ and $\rho^{0}$ modes are modelled with Crystal-Ball function [10] whereas Crystal Ball with two Gaussians is used for the $\overline{K}^{*0}$ mode. To account for the data-MC difference, a free offset and scale factor are implemented for the mean and width of the $\overline{K}^{*0}$ PDF, and the obtained values are used for the other two signal modes. The $\pi^{0}$ and $\eta$ $M_{D^{0}}$ background distributions are modelled in one of two ways: (1) a single Crystal Ball, or (2) combinations of either Crystal Ball or logarithmic Gaussians [11] with upto two Gaussians. There is only one $\pi^{0}$ type background in the $\phi$ mode, which is $D^{0}\rightarrow \phi\pi^{0}$. The $\pi^{0}$ and $\eta$ type backgrounds in the $\overline{K}^{*0}$ modes are: $D^{0}\rightarrow \overline{K}^{*0} \pi^{0}$, $K^{-}\rho^{+}$, $K_{0}^{*}(1430)^{-}\pi^{+}$, $K^{*-}\pi^{+}$, non-resonant $K^{-}\pi^{+}\pi^{0}$, $\overline{K}^{*0}\eta$ and non-resonant $K^{-}\pi^{+}\eta$. The backgrounds in the $\rho^{0}$ mode are $\rho^{0}\pi^{0}$, $\rho\pi$ and $K^{-}\rho^{+}$ due to the misidentification of a kaon as a pion. In all of the three modes, apart from the above mentioned backgrounds, another category is defined for all other decays with correctly reconstructed $D^{0}$. Two more backgrounds are present in the $\overline{K}^{*0}$ mode: non-resonant decay to $K^{-}\pi^{+}(\pi^{+}\pi^{-})$ with the photon being emitted as FSR, and the decay to $K^{-}\rho^{+}$ with the photon being emitted from a radiative decay of the charged $\rho$ meson. The $M_{D^{0}}$ distributions are similar to that of signal in these modes as there are no missing particles, which are fixed from MC and known branching fractions. The remaining combinatoric backgrounds in $M_{D^{0}}$ is modelled with an exponential function in the $\phi$ mode and a second order Chebyshev polynomial in the $\overline{K}^{*0}$ and $\rho^{0}$ modes.  

~~~The $M_{D^{0}}$ distribution for the dominant background is calibrated by reconstructing $D^{0}\rightarrow K_{S}^{0}\gamma$ in both simulation and data. Since the decay is forbidden, this yields mostly the backgrounds of the type $D^{0}\rightarrow K_{S}^{0}\pi^{0}$ and $D^{0}\rightarrow K_{S}^{0}\eta$. The same selection cuts as in the $\phi$ mode are applied here. $K_{S}^{0}$ candidates have a mass cut of $\pm$9 MeV/c$^{2}$ around the nominal $K_{S}^{0}$ mass. 

~~~The $\cos{\theta_{H}}$ distribution is modelled with 1$-\cos^{2}{\theta_{H}}$, where all the parameters are fixed to MC values for all the three signal modes. The $V\pi^{0}$ and $V\eta$ categories are parametrized by $\cos^{2}{\theta_{H}}$ for the $\phi$ mode, a third order Chebyshev polynomial for the $\overline{K}^{*0}$ mode and a second order polynomial for the $\rho^{0}$ mode. The other background categories are parametrized with respect to their shapes. The fit results are shown in Fig. 1, Fig. 2 and Fig. 3 for the $\phi$, $\overline{K}^{*0}$ and $\rho^{0}$, respectively. The signal component is shown by dotted red line. The signal yield obtained from the fits are: 524 $\pm$ 35 for $\phi$ mode, 9104 $\pm$ 396 for $\overline{K}^{*0}$ mode and 500 $\pm$ 85 for $\rho^{0}$ mode, respectively. Thus the raw asymmetries are $-$0.091 $\pm$ 0.066 (stat), $-$0.002 $\pm$ 0.020 (stat), 0.056 $\pm$ 0.151 (stat) for $\phi$, $\overline{K}^{*0}$ and $\rho^{0}$ modes, respectively.

\begin{figure}[ht!] \centering \begin{tabular}{cccc} \includegraphics[width=0.22\columnwidth]{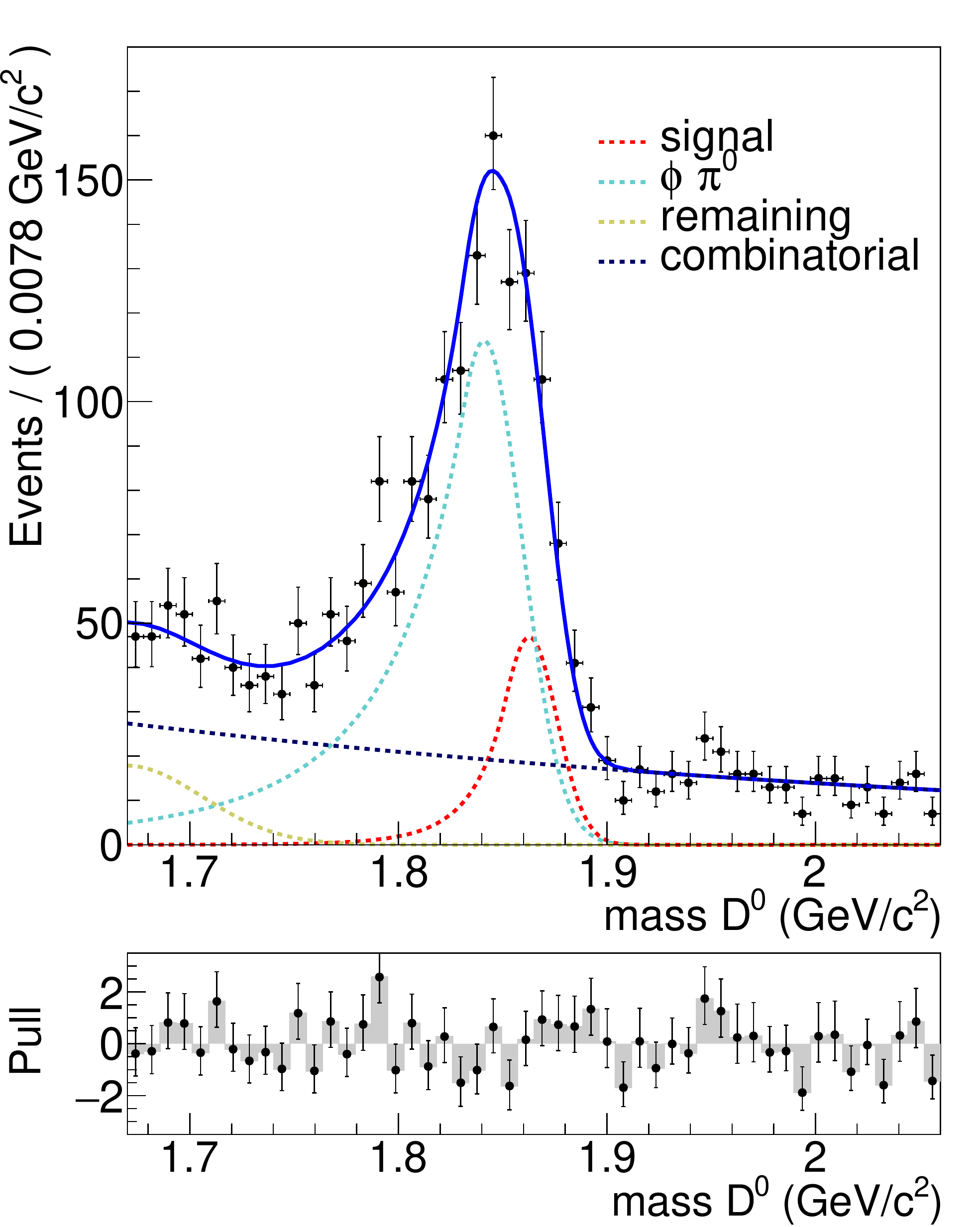}
& \includegraphics[width=0.22\columnwidth]{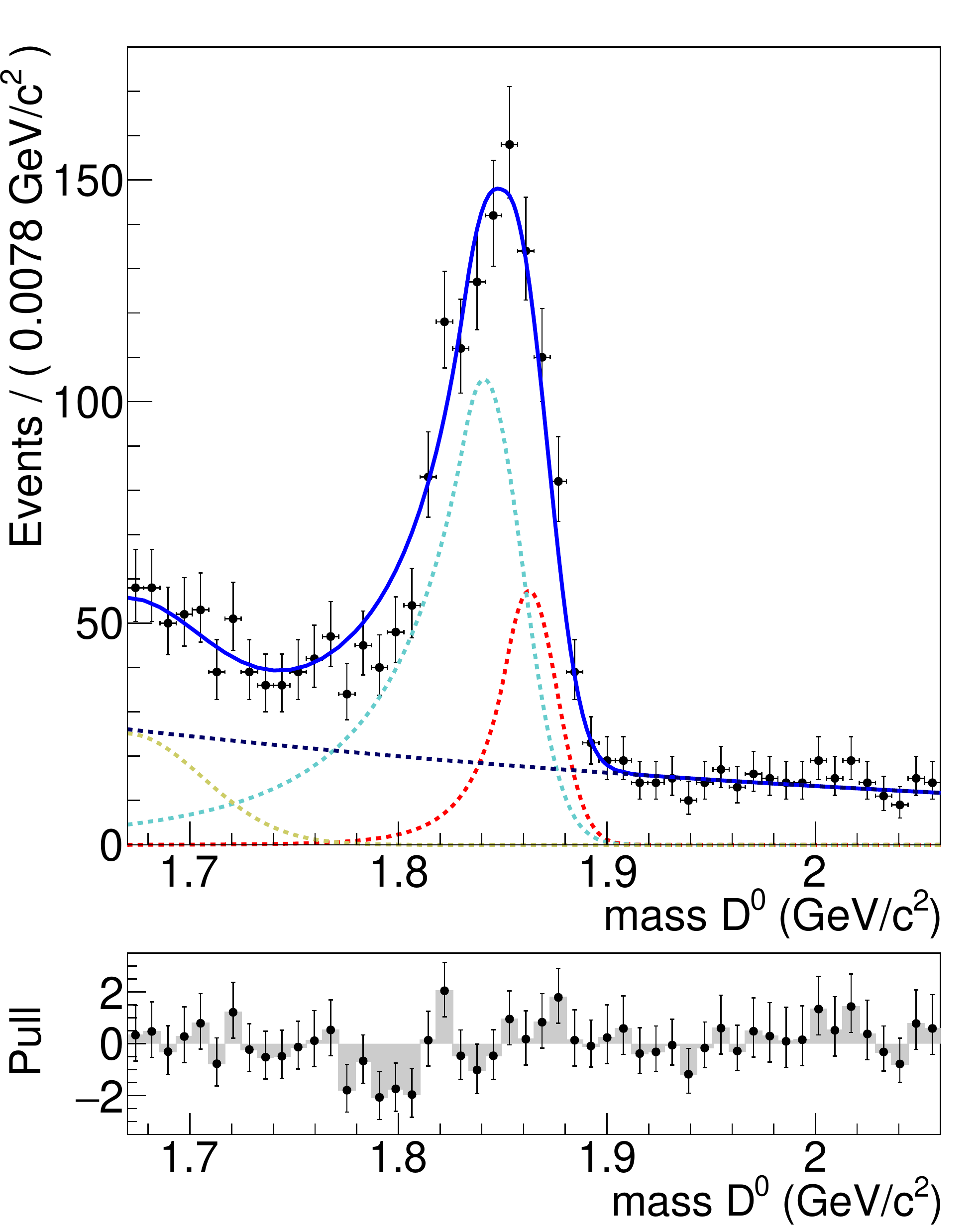}
&\includegraphics[width=0.22\columnwidth]{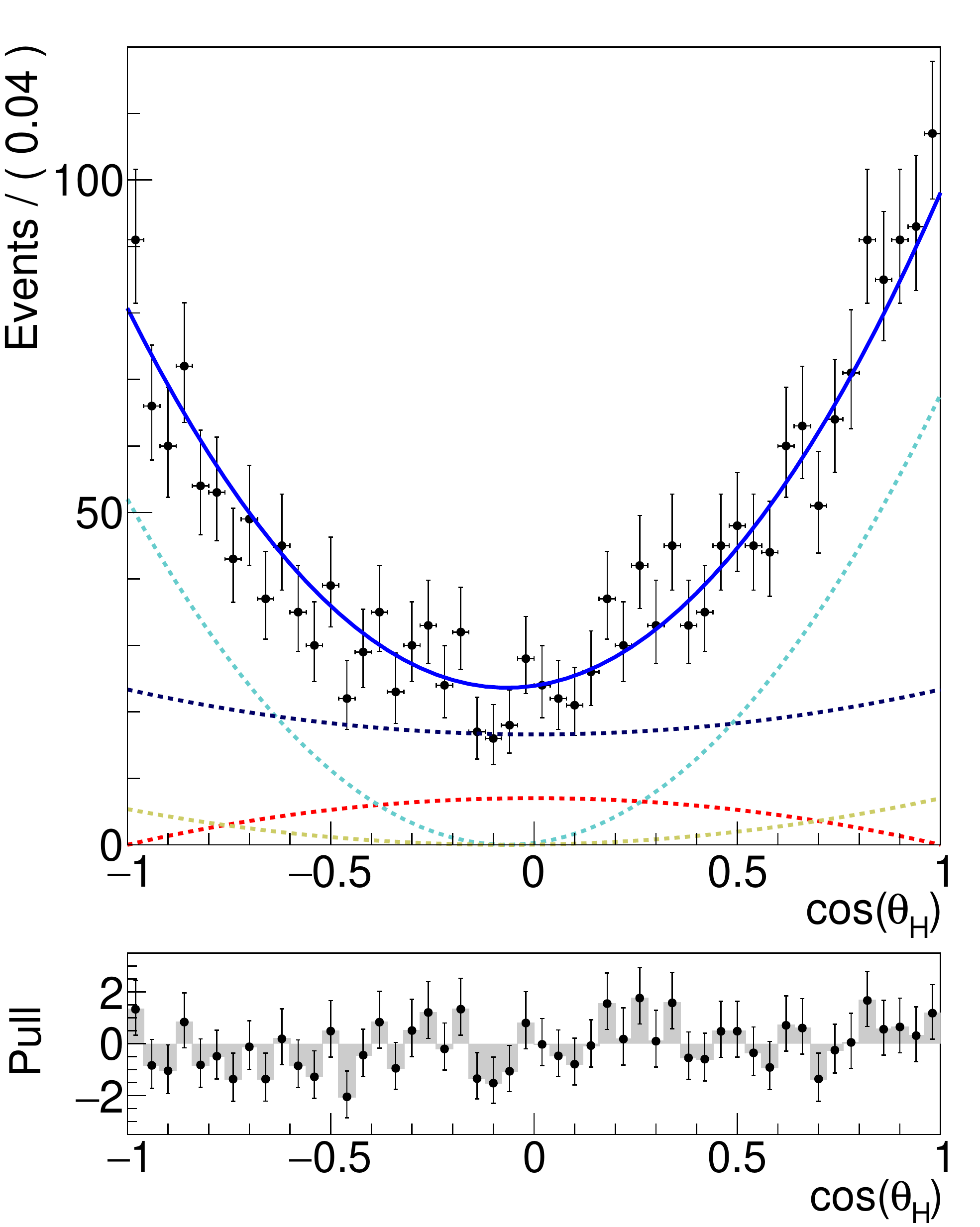}
& \includegraphics[width=0.22\columnwidth]{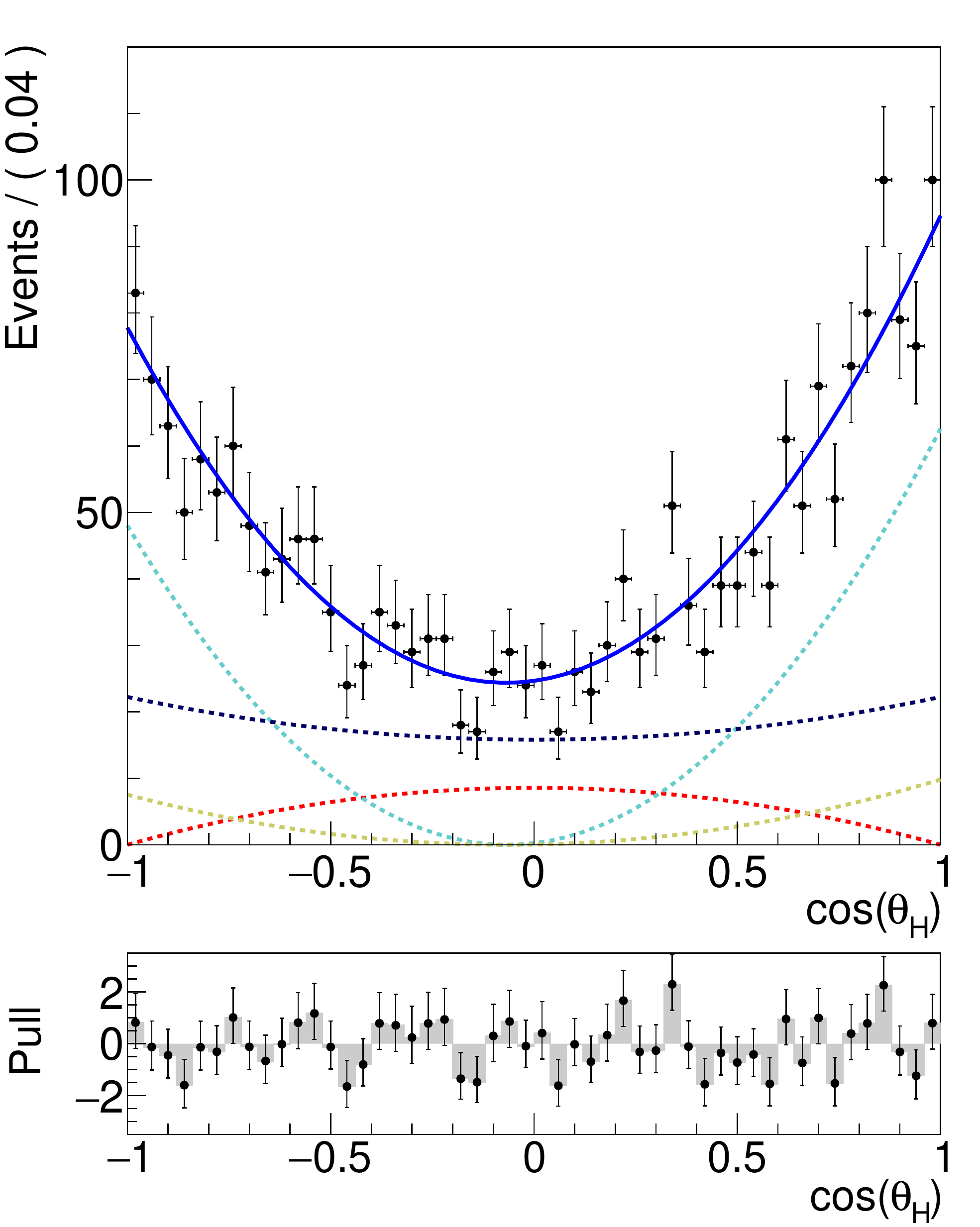}\\

\end{tabular}\caption{$M_{D^{0}}$ distributions for $D^{0}$ (first) and $\overline{D^{0}}$ (second) and $\cos{\theta_{H}}$ distributions for $D^{0}$ (third) and $\overline{D^{0}}$ (fourth) in the $\phi$ mode. The dotted red, cyan, blue and yellow lines are signal, $\phi\pi^{0}$, combinatoric and remaining components, respectively.}
\end{figure}

\begin{figure}[ht!] \centering \begin{tabular}{cccc} \includegraphics[width=0.22\columnwidth]{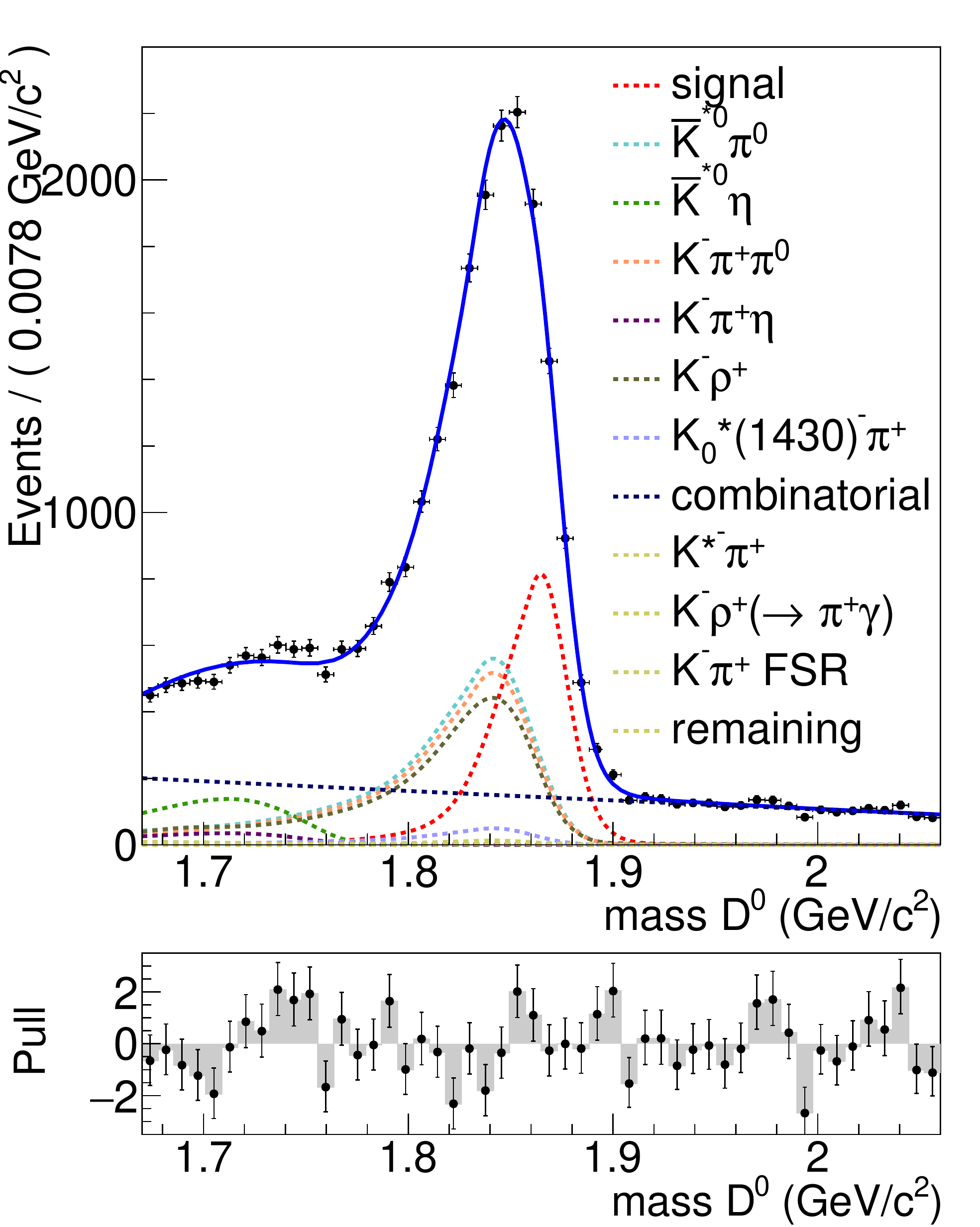}
& \includegraphics[width=0.22\columnwidth]{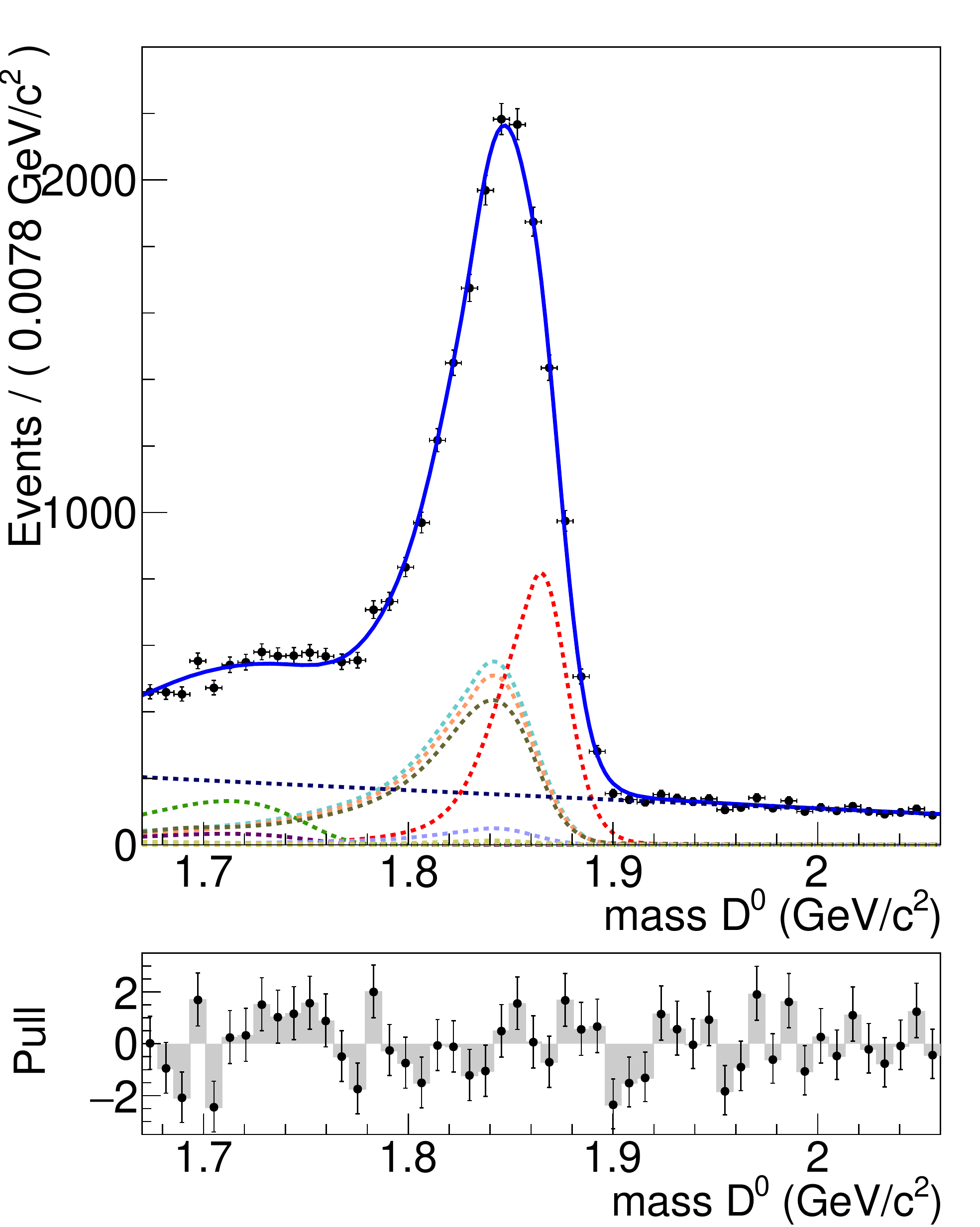}
&\includegraphics[width=0.22\columnwidth]{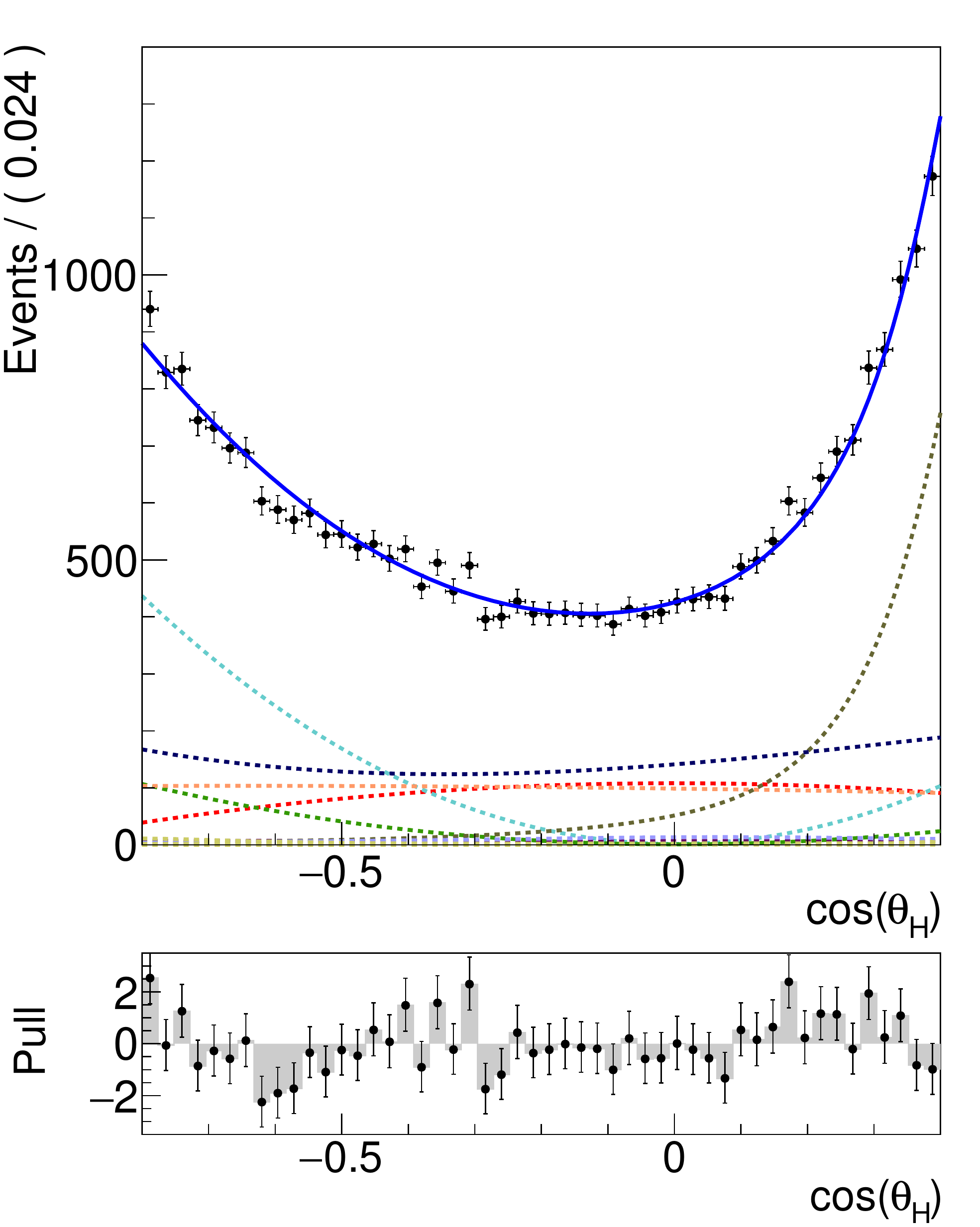}
& \includegraphics[width=0.22\columnwidth]{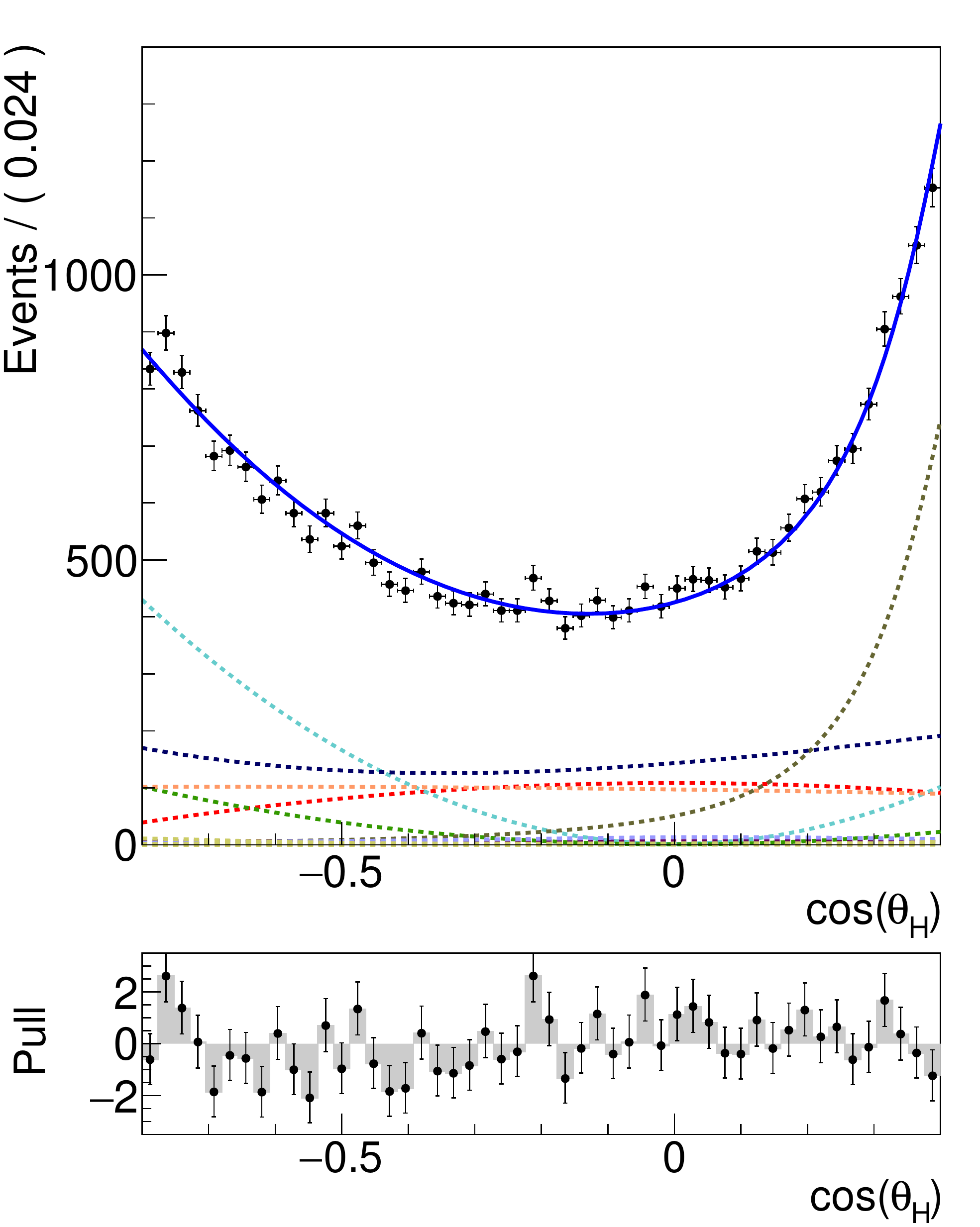}\\

\end{tabular}\caption{$M_{D^{0}}$ distributions for $D^{0}$ (first) and $\overline{D^{0}}$ (second) and $\cos{\theta_{H}}$ distributions for $D^{0}$ (third) and $\overline{D^{0}}$ (fourth) in the $\overline{K}^{*0}$ mode. The dotted red, blue lines are the signal and combinatoric components, respectively. The other components are also shown in different colours.}
\end{figure}

\begin{figure}[ht!] \centering \begin{tabular}{cccc} \includegraphics[width=0.22\columnwidth]{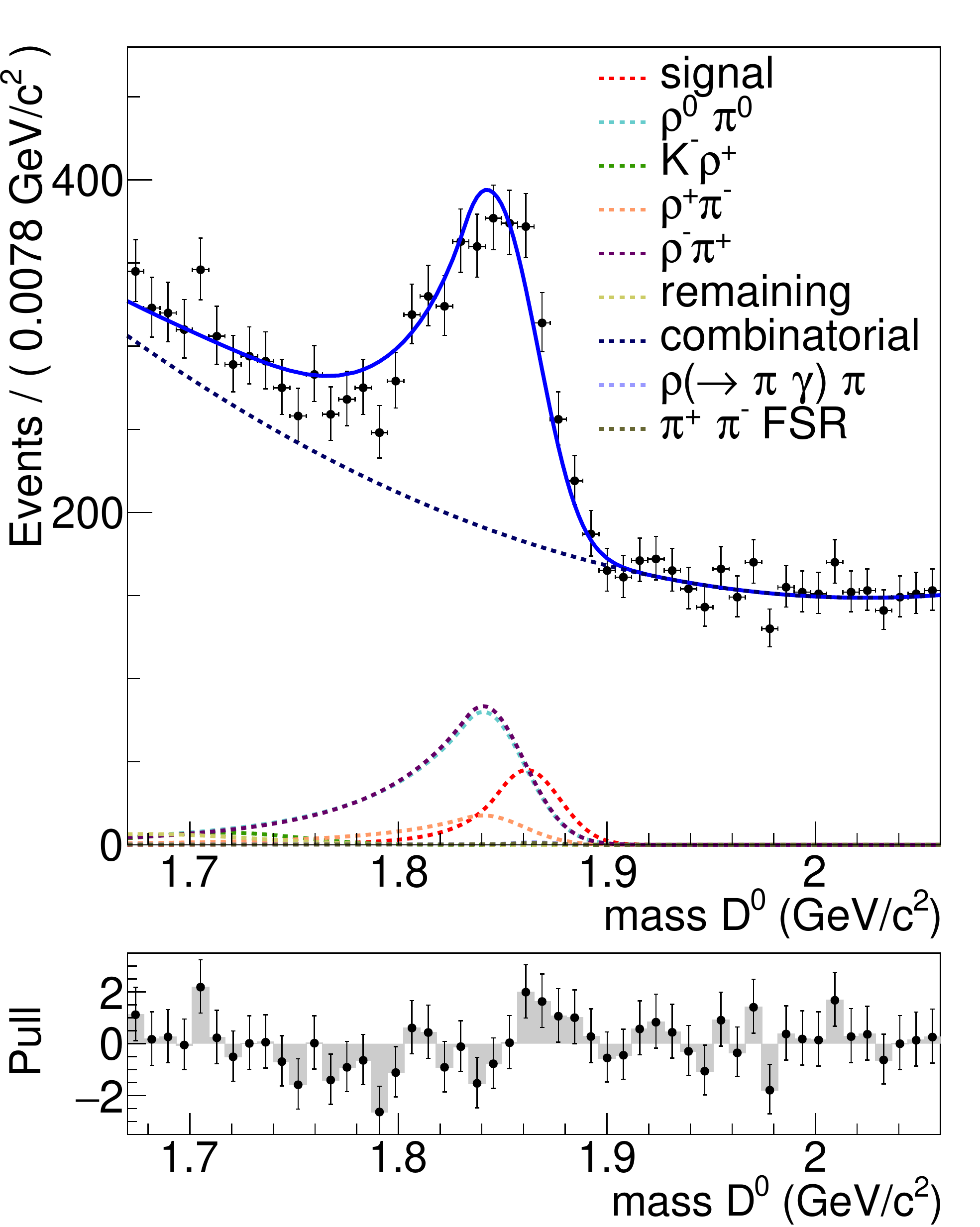}
& \includegraphics[width=0.22\columnwidth]{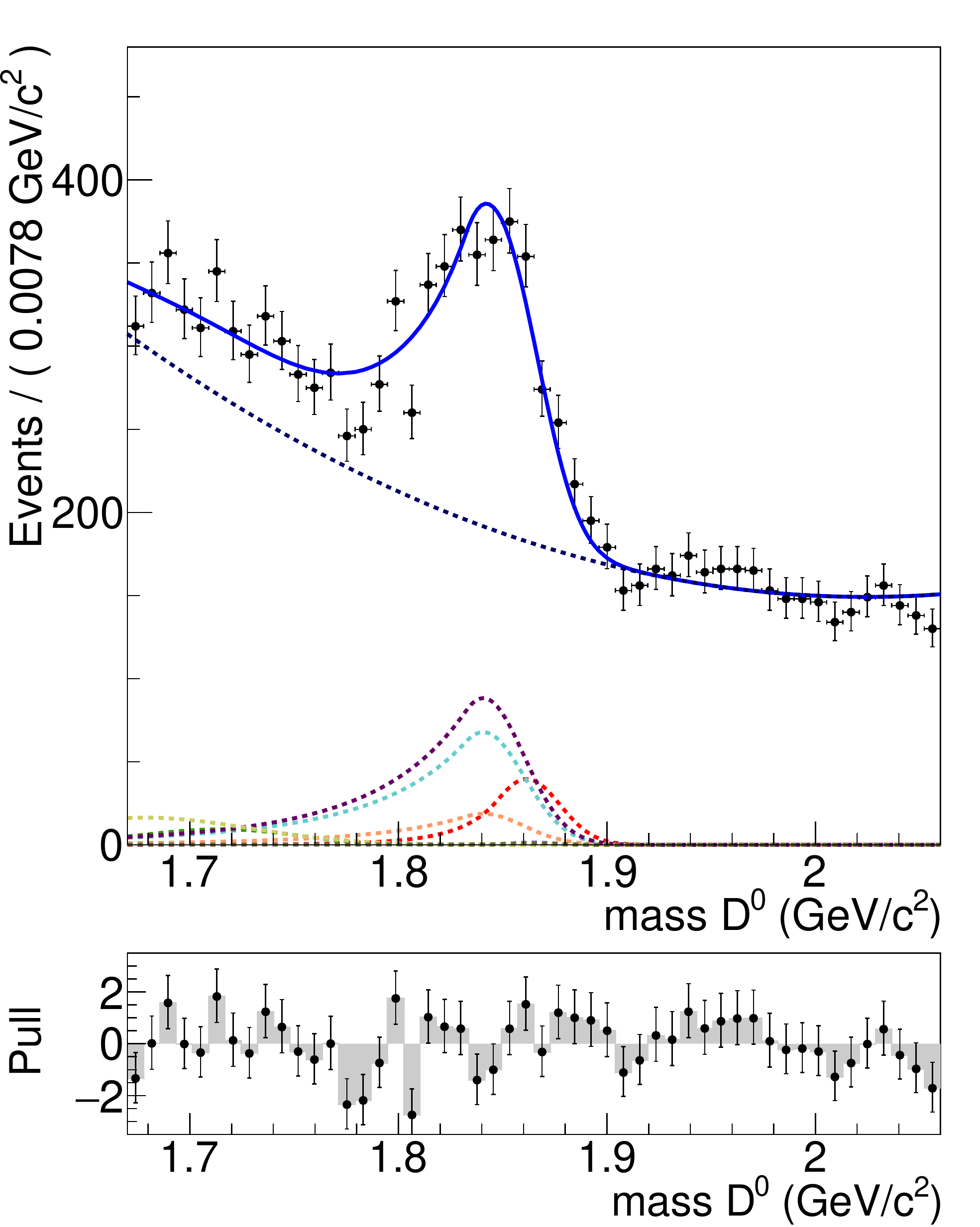}
&\includegraphics[width=0.22\columnwidth]{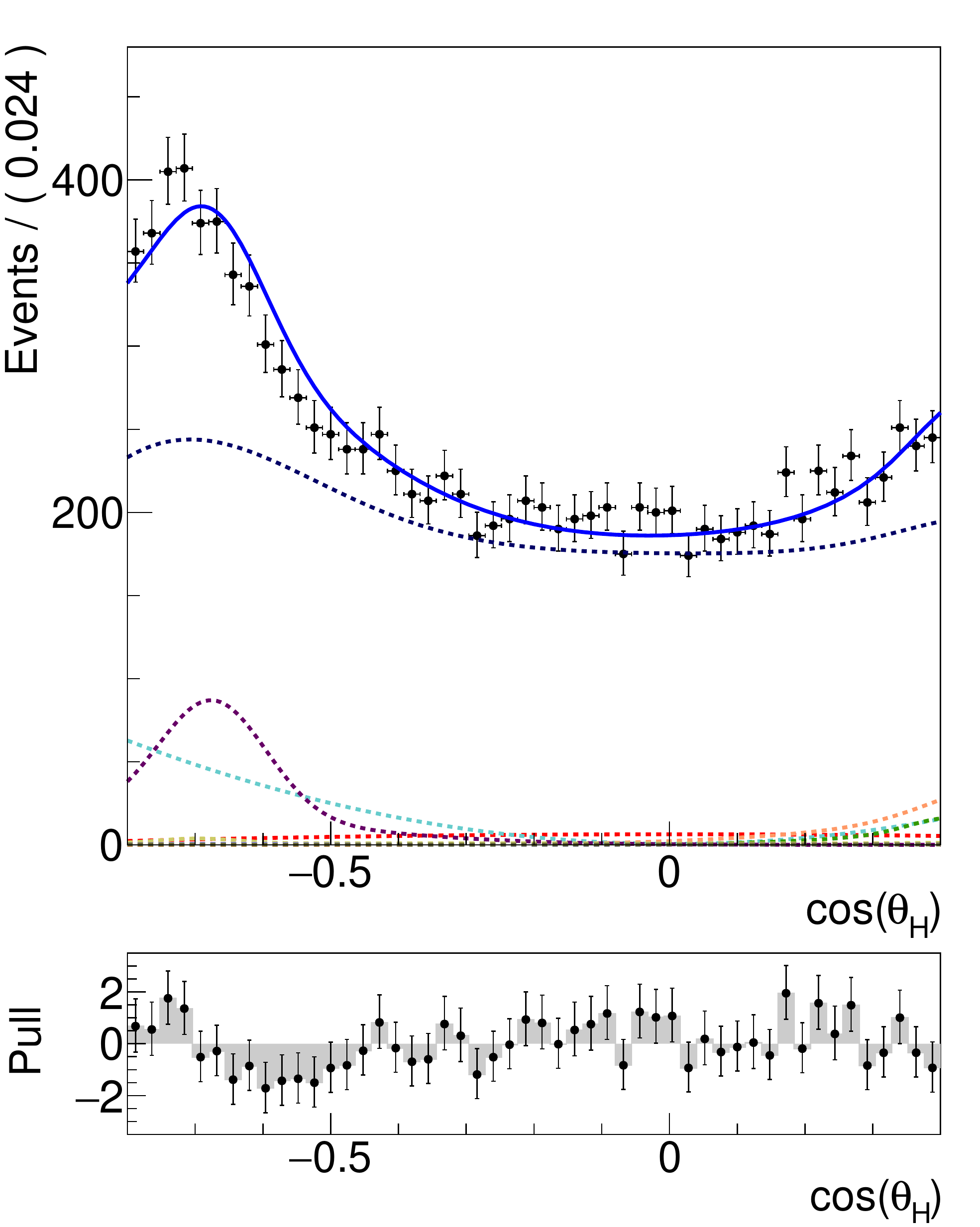}
& \includegraphics[width=0.22\columnwidth]{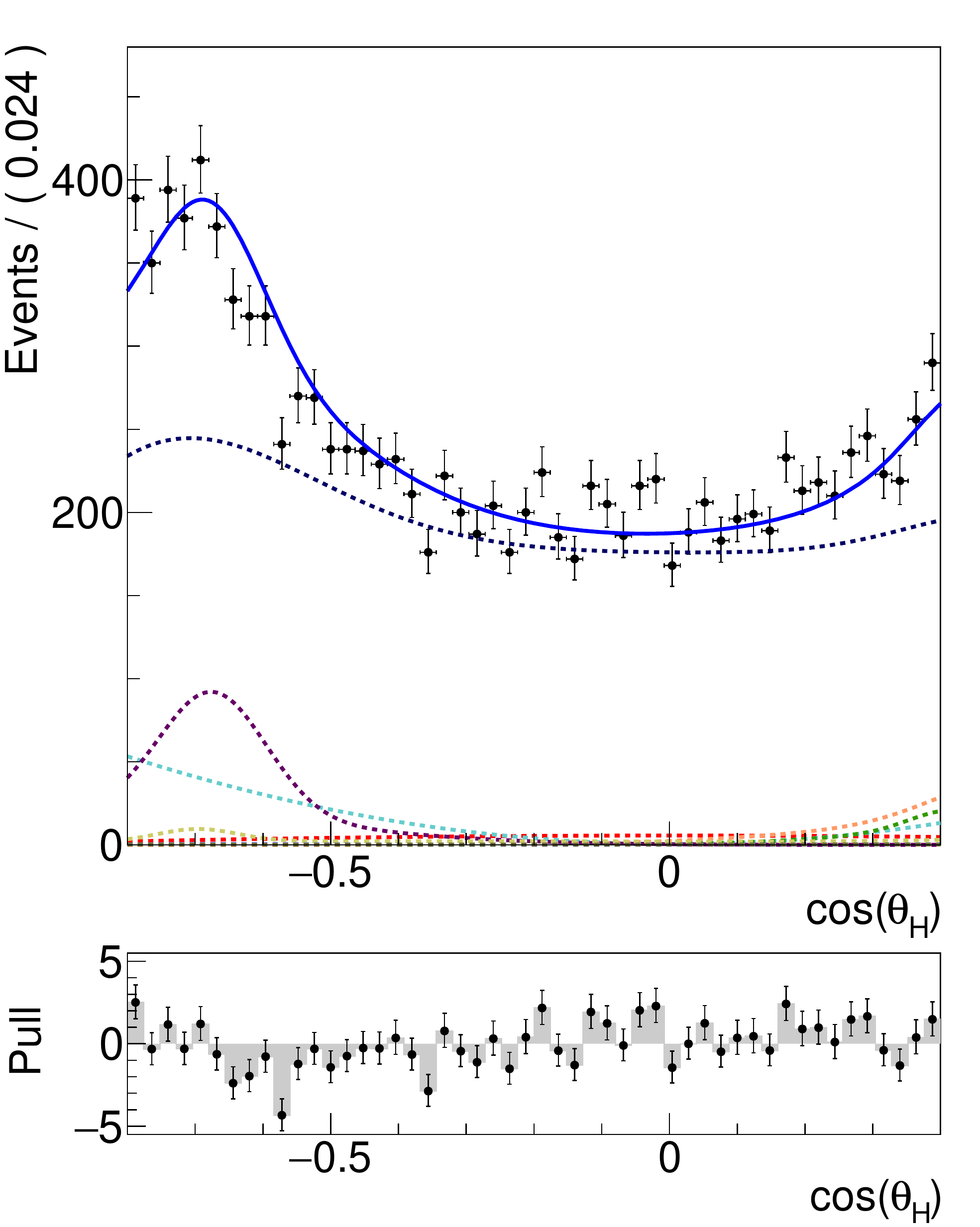}\\

\end{tabular}\caption{$M_{D^{0}}$ distributions for $D^{0}$ (first) and $\overline{D^{0}}$ (second) and $\cos{\theta_{H}}$ distributions for $D^{0}$ (third) and $\overline{D^{0}}$ (fourth) in the $\rho^{0}$ mode. The dotted red, blue lines are the signal and combinatoric components, respectively. The other components are also shown in different colours.}
\end{figure}

~~~The control samples are analysed in the similar way as done earlier by the Belle collaboration [12]. The selection criteria are the same as in the signal modes and the signal extraction is performed via sideband subtraction method, where the signal (SW), lower (LW) and upper (UW) windows are defined in $M_{D^{0}}$. The efficiencies from the simulation is obtained as 22.7$\%$, 27.0$\%$ and 21.4$\%$ for $K^{+}K^{-}$, $K^{-}\pi^{+}$ and $\pi^{+}\pi^{-}$ samples respectively. The extracted signal yields are 362,274 for $K^{+}K^{-}$ mode, 4.02 $\times$ 10$^{6}$ for $K^{-}\pi^{+}$ mode and 127,683 for $\pi^{+}\pi^{-}$ mode. The raw asymmetries are (2.2 $\pm$ 1.7) $\times$ 10$^{-3}$ for $K^{+}K^{-}$ mode, (1.3 $\pm$ 0.5) $\times$ 10$^{-3}$ for $K^{-}\pi^{+}$ mode and (8.1 $\pm$ 3.0) $\times$ 10$^{-3}$ for $\pi^{+}\pi^{-}$ mode.

\subsection{Systematics}
\label{sec-2}
~~~The list of various sources of systematics is shown in Table 2. There are two main sources: one is due to the applied selection criteria and the other due to the signal extraction method for both the signal and the control sample modes. The uncertainties common to both signal and normalization modes cancel. A 2$\%$ uncertainty is given to the photon reconstruction efficiency [13]. The systematic effect from the variable $q$ is due to its low resolution in the signal mode compared to the control sample mode. This is evaluated with the help of the control sample $D^{0}\rightarrow \overline{K}^{*0}\pi^{0}$. The systematic uncertainty for the $q$ cut is found to be 1.16$\%$. The systematic for the $\pi^{0}$ veto is calculated with the mode $D^{0}\rightarrow K_{S}^{0} \pi^{0}$. The veto is performed with the first daughter of $\pi^{0}$ with all other photons except the second $\pi^{0}$ daughter. The ratio R between the yields for the signal and the control sample is calculated for both simulation and the data and by taking the double ratio, $\frac{R_{\mathrm{MC}}}{R_{\mathrm{DATA}}}$ we assigned the systematic for the $\pi^{0}$ veto. Similarly we extracted the systematics for the $\frac{E_{9}}{E_{25}}$ variable with $D^{0}\rightarrow K_{S}^{0}\gamma$ as the control sample. A systematic uncertainty is assigned to the fit procedure where the fixed parameters are varied with respect to their errors. The biggest difference between the so obtained $\mathcal{B}$ and $A_{CP}$ and the mean value is taken as the systematic error. An uncertainty is assigned to the dominant backgrounds of $\pi^{0}$ type for the chosen width of the smearing, varying the width by $\pm$1 MeV/c$^{2}$. For the normalization mode systematics, the procedure is repeated with different sidebands, $\pm$25 MeV/c$^{2}$ for $M_{D^{0}}$. The statistical error on the fraction $f$, which is the fraction of background events in the signal window compared to all events in the sideband, is also taken into account. The difference between the data and simulation could also affect $f$, so a similar procedure for the calibration of the $\pi^{0}$ background is also performed. A systematic uncertainty is assigned for the case when $f$ is obtained from simulation smeared by a Gaussian of width 1.6 MeV. 
\begin{table}[ht!]
\footnotesize \centering \caption{List of various sources of systematics and their contributions for $\mathcal{B}$ and $A_{CP}$ in $D^{0}\rightarrow V \gamma$ study.}
\begin{tabular}{l|l|l|l}

  \hline 
 Source &~~~~~~~	$D^{0}\rightarrow \phi \gamma$		& ~~~~~~~$D^{0}\rightarrow \overline{K}^{*0} \gamma$		&	~~~~~~~$D^{0}\rightarrow \rho \gamma$	 \\
  \hline
  &	$\mathcal{B}$ ($\%$) \hspace{0.1 in} $A_{CP} \times 10^{-3}$ & $\mathcal{B}$ ($\%$) \hspace{0.1 in} $A_{CP} \times 10^{-3}$ 	 & $\mathcal{B}$ ($\%$) \hspace{0.1 in} $A_{CP} \times 10^{-3}$	\\
  \hline
  $\gamma$ rec. eff &	 2 \hspace{0.562 in}	$-$	& 2	\hspace{0.585 in} $-$	& 2	\hspace{0.595 in}$-$ \\
  $\Delta M$ 		&	1.16	 \hspace{0.422 in}	$-$		& 1.16 \hspace{0.445 in}	$-$		& 1.16	\hspace{0.43 in}	$-$	\\
  $\pi^{0}$ veto		& 0.5\hspace{0.507 in}	$-$		&0.5		\hspace{0.5 in}	$-$	&0.5\hspace{0.5175 in}	$-$ \\
  $E_{9}/E_{25}$	&	0.96	\hspace{0.422 in}	$-$	&		0.96	 \hspace{0.445 in}	$-$ 	& 0.96 \hspace{0.435 in}	$-$ \\
  Signal shape 	&	1.39	\hspace{0.37 in}	 0.32		&  ~$-$	\hspace{0.54 in}	$-$		&2.33	\hspace{0.345 in}		 4.29 \\
  Background & 0.95	\hspace{0.37 in}	 0.30		&  ~2.81	\hspace{0.33 in}	0.41		&3.00	\hspace{0.345 in}		 3.78 \\
   shape	&		&		&\\
  Norm modes	 & 0.05	\hspace{0.37 in}	 0.46		&  ~0.00	\hspace{0.33 in}	0.01		&0.14	\hspace{0.345 in}		 0.54 \\
  systematics &					&			&	\\
  \hline	
  Total		&	3.06	\hspace{0.37 in}	 0.64		&  ~3.80	\hspace{0.33 in}	0.41		&4.58	\hspace{0.345 in}		 5.74 \\
  \hline
 
  \end{tabular}

  \end{table}

\subsection{Results}
\label{sec-2}
~~~The preliminary measurements of the branching fraction in $D^{0}\rightarrow V\gamma$ mode are

\begin{align*}
\mathcal{B}(D^{0}\rightarrow \phi \gamma) &= (2.76 \pm 0.20 \pm 0.08) \times 10^{-5}, \\
\mathcal{B}(D^{0}\rightarrow \overline{K}^{*0} \gamma) &= (4.66 \pm 0.21 \pm 0.18) \times 10^{-4}, \\
\mathcal{B}(D^{0}\rightarrow \rho^{0} \gamma) &= (1.77 \pm 0.30 \pm 0.08) \times 10^{-5},\\
 \end{align*}
where the uncertainties are statistical and systematic, respectively. The branching fraction for the $\phi$ mode is consistent with the previous measurements. There is a 3.3$\sigma$ difference with the BABAR result for the branching fraction of $D^{0}\rightarrow \overline{K}^{*0} \gamma$ mode. We also report the first branching fraction measurement for the  $D^{0}\rightarrow \rho^{0}\gamma$ mode with $> 5\sigma$ significance.  The observed value of $\mathcal{B}$ for the $\rho^{0}$ mode is very close to that of $\phi$ mode, agreeing with the theory predictions. We also measure the first-ever $A_{CP}$ measurements for $D^{0}\rightarrow V\gamma	$ modes. The preliminary results are 

\begin{align*}
A_{CP}(D^{0}\rightarrow \phi \gamma) &= -0.094 \pm 0.066 \pm 0.001,  \\
A_{CP} (D^{0}\rightarrow \overline{K}^{*0} \gamma) &= -0.003 \pm 0.020 \pm 0.000, \\
A_{CP} (D^{0}\rightarrow \rho^{0} \gamma) &= \phantom{-}0.056 \pm 0.151 \pm 0.006, 
\end{align*}
where the uncertainties are statistical and systematic, respectively. No $CP$ asymmetry is observed.

\section{$D^{0}\rightarrow \gamma \gamma$ decay}
\label{sec-1} 
 
~~~Flavour changing neutral currents (FCNC) in the Standard Model are forbidden at the tree level of an interaction, while it can happen through higher orders. There are many new physics models that allow FCNC even at the tree level by a $c \rightarrow u \gamma \gamma$ transition. The branching fraction is expected to be very small ($\mathcal{O}(10^{-8})$) [14], [15], [16]. But with the minimal supersymmetric Standard Model (MSSM), it has been predicted that the branching fraction can be enhanced due to the exchange of a gluino to $\mathcal{O}(10^{-6})$ [17]. Thus by measuring the mode $D^{0}\rightarrow \gamma \gamma$, we could identify any new physics contributions [18].

~~~Previous measurements were carried out by BABAR [19], CLEO [20] and BESIII [21] collaborations. The most stringent limit is set by the BABAR collaboration: 2.2$\times$10$^{-6}$ with a confidence level of 90$\%$. Our analysis is with the 832 fb$^{-1}$ data collected at the $\Upsilon(4S)$ and $\Upsilon(5S)$ resonances. As in the $D^{0}\rightarrow V\gamma$ analysis, we also performed a Monte Carlo simulation for the selection criteria and background studies. 

\subsection{Selection criteria, fitting and signal extraction}
\label{sec-2}

~~~The analysis is similar to that of $D^{0}\rightarrow V\gamma$ decay. We use the mode $D^{*}\rightarrow D^{0}\pi_{\mathrm{slow}}$ to suppress the large combinatoric background. But there are peaking backgrounds due to the decays of a $\pi^{0}$ and/or $\eta$ meson, which decays to a pair of photons. These are $D^{0}\rightarrow \pi^{0} \pi^{0}$, $D^{0}\rightarrow \eta \pi^{0}$, $D^{0}\rightarrow \eta \eta$, $D^{0}\rightarrow K_{S}^{0}\pi^{0}$ and $D^{0}\rightarrow K_{L}^{0} \pi^{0}$. To suppress these peaking backgrounds, a dedicated $\pi^{0}(\eta)$ veto is applied and the suppression of merged clusters in ECL by $E_{9}/E_{25}$. The chosen control sample mode is $D^{0}\rightarrow K_{S}^{0} \pi^{0}$ to measure the branching fraction. 

~~~The signal extraction is performed with a two-dimensional fit between $M_{D^{0}}$ and $\Delta M$ variables. The efficiency is found to be 7.3$\%$. We obtained a signal yield 4 $\pm$ 15 from the fit. The fit is shown in Fig. 4. The similar procedure is done for the control sample mode and we obtained a signal yield 343050 $\pm$ 673.

\begin{figure}[h]
\centering
\includegraphics[width=7cm,clip]{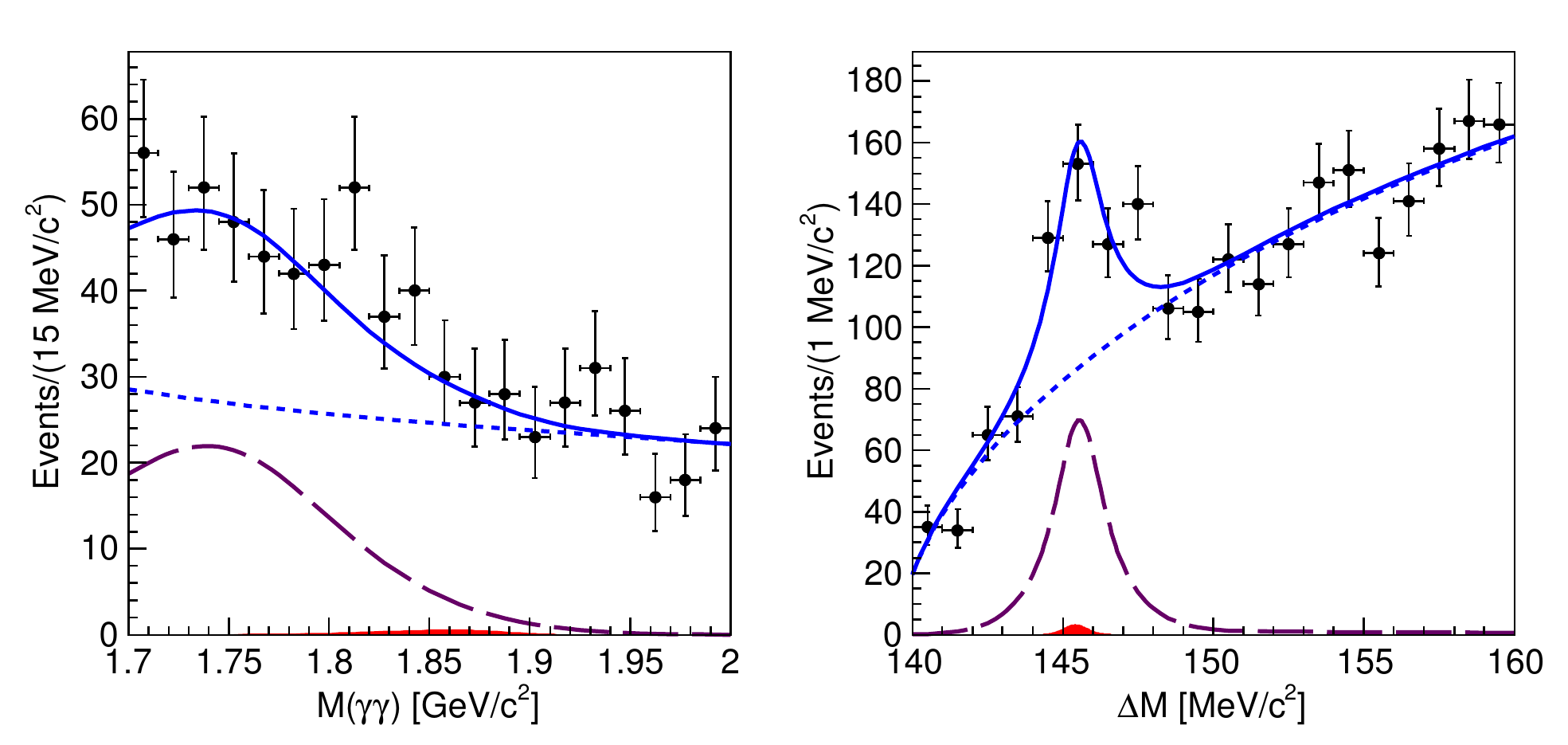}
\caption{2D fit between $\Delta M$ (left) and $M_{D^{0}}$ (right) where the dashed blue, purple lines are combinatoric, peaking backgrounds respectively and the red histogram is the signal.}
\end{figure}

\subsection{Systematics}
\label{sec-2}

~~~The dominant contribution is coming from the cut variation of $E_{\gamma2}$, $A_{E}$ and $\mathcal{P}(\pi^{0})$, which are the energy of lower energy photon, energy asymmetry between the two photons and the probability of $\pi^{0}$, respectively. For $E_{\gamma2}$,  we calculated $\frac{N}{\epsilon}$, where $N$ is signal yield and $\epsilon$ is the detection efficiency, with and without any photon requirement on the photon energy in the $D^{0}\rightarrow \phi\gamma$ control sample. The change with respect to the nominal value is taken as the systematic error. $\mathcal{P}(\pi^{0})$ systematics is estimated with the same control sample where we relaxed the selection to $\mathcal{P}(\pi^{0}) < 0.7$ from $\mathcal{P}(\pi^{0}) < 0.15$. We double the above two systematic uncertainties as we have two photons in the final state. For $A_{E}$, as we do not have any proper control sample, we fit to the data without any requirement on $A_{E}$ and take the resulting change in the upper limit as the systematic error. The list of all the sources of systematics are shown in Table 3. 

\begin{table}[ht!]
\centering
\begin{tabular}{l|l}
  \hline 
Source	&	Contribution \\
\hline
cut variation	&	$\pm$ 6.8 $\%$\\
signal shape 	&	$^{+4.0}_{-2.4}$ events \\
$\gamma$ rec. eff & 	$\pm$ 4.4 $\%$ \\
$K_{S}^{0}$ reconstruction &	$\pm$ 0.7 $\%$ \\
$\pi^{0}$ identification	&	$\pm$ 4.0 $\%$ \\
$\mathcal{B}(D^{0}\rightarrow K_{S}^{0} \pi^{0})$ &	$\pm$ 3.3 $\%$ \\
\hline
  \end{tabular}
  
  \caption{Systematic uncertainties for $D^{0}\rightarrow \gamma \gamma$ study.}
  \end{table}

\subsection{Results}
\label{sec-2}

~~~As the signal is absent in this analysis, a frequentist method is used to set the upper limit on the branching fraction of $D^{0}\rightarrow \gamma \gamma$ decay with 90$\%$ confidence level. The result is
\begin{equation}
			\mathcal{B}(D^{0}\rightarrow \gamma \gamma) < 8.4 \times 10^{-7}.
\end{equation}
This is the most stringent limit of this mode to date, as illustrated in Fig. 5. 

\begin{figure}[h]
\centering
\includegraphics[width=7cm,clip]{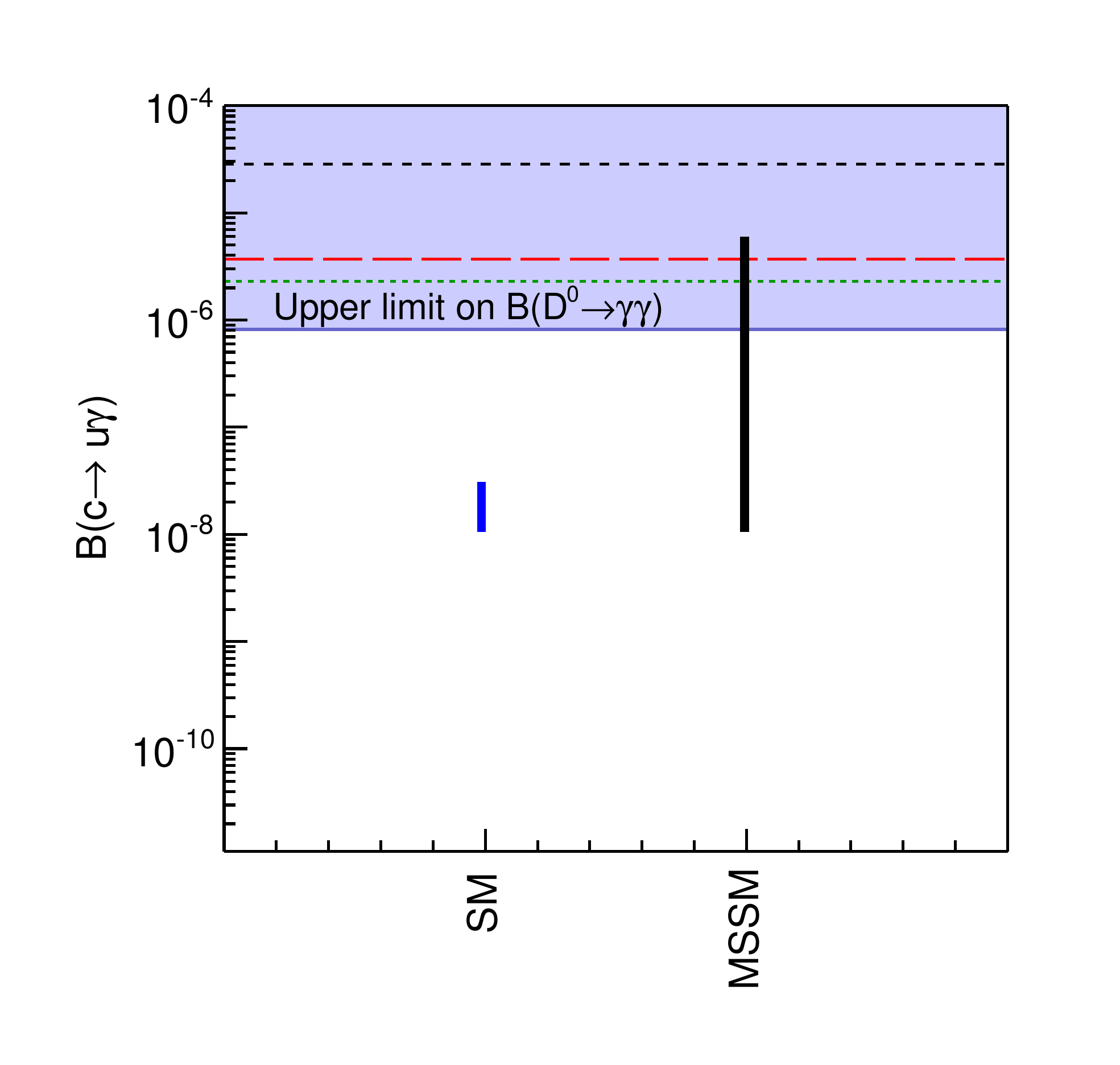}
\caption{$\mathcal{B}(c\rightarrow u\gamma)$ prediction in SM and MSSM. Our result is shown in purple line. Dotted green, red and black lines are the results from BABAR, BESIII and CLEO collaborations, respectively.}
\end{figure}

\section{Acknowledgements}
\label{sec-1} 

~~~We thank the KEKB group and all institutes and agencies that support the work of the members of the Belle collaboration. We also extend our gratitude to Indian Institute of Technology Madras (IIT Madras) and International and Alumni relations, IIT Madras for their financial support to attend the conference.

%
%
%


\end{document}